\documentclass[aps,prd,amsmath,amssymb,superscriptaddress,nofootinbib,10pt,twocolumn]{revtex4-2}
\usepackage{amsmath,amssymb,bm,color,float,graphicx,mathrsfs}
\usepackage[hidelinks]{hyperref}
\hypersetup{colorlinks=true,linkcolor=blue,citecolor=blue,urlcolor=blue}
\usepackage{Macro}
\usepackage{setspace}
\usepackage{soul}
\setstcolor{red}
\usepackage{graphicx}	
\usepackage{amsmath}	
\usepackage{diagbox}
\usepackage{multirow}
\usepackage{array}
\usepackage{verbatim}
\newcolumntype{M}[1]{>{\centering\arraybackslash}m{#1}}
\usepackage{amsmath}	

\definecolor{mygreen}{rgb}{0,0.5,0}  

\newcommand{\scaledtable}[2][\textwidth]{\resizebox{#1}{!}{#2}}

\newcommand{\compile}{}

\def\fnl{f_{\rm NL}}

\begin{document}

\ifdefined\compile
\title{Constraining primordial non-Gaussianity and energy injection with the thermal Sunyaev-Zeldovich effect and integrated Sachs-Wolfe effect cross-correlation}

\author{Ayodeji Ibitoye}\thanks{Corresponding author:  A.Ibitoye, \url{ayodeji.ibitoye@gtiit.edu.cn}}
\affiliation{Department of Physics, Guangdong Technion - Israel Institute of Technology, Shantou, Guangdong 515063, P.R. China}
\affiliation{Centre for Space Research, North-West University, Potchefstroom 2520, South Africa}
\affiliation{Department of Physics and Electronics, Adekunle Ajasin University, P. M. B. 001, Akungba-Akoko, Ondo State, Nigeria}
\author{Yin-Zhe Ma}\thanks{Corresponding author: Y.-Z. Ma, \url{mayinzhe@sun.ac.za}}
\affiliation{Department of Physics, Stellenbosch University, Matieland 7602, South Africa}
\affiliation{National Institute for Theoretical and Computational Sciences (NITheCS) South Africa}
\author{Prabhakar Tiwari}
\affiliation{Department of Physics, Guangdong Technion - Israel Institute of Technology, Shantou, Guangdong 515063, P.R. China}

\date{\today}

\begin{abstract}
Constraining primordial non-Gaussianity (PNG) provides key insights
into the physics of cosmic inflation and the initial conditions of
the Universe, which remain central topics in cosmology. In this
study, we use the cross-correlation between the integrated
Sachs-Wolfe (ISW) effect and the thermal Sunyaev-Zeldovich (tSZ)
effect derived from Ibitoye et al.~\cite{Ibitoye24} to jointly constrain PNG
and early-Universe energy injection, including the standard
intergalactic medium contribution. For scale-independent PNG we
obtain {$f_{\rm NL} = -358^{+140}_{-114}$}
($68\%$~C.L.). For a scale-dependent model
($f_{\rm NL}=f_{\rm NL}^{0}(\ell/\ell_{0})^{n_{\rm NL}}$, with
$\ell_{0}=200$), we find
{$f^{0}_{\rm NL} = -296^{+173}_{-157}$} and
{$n_{\rm NL} = 0.62^{+1.02}_{-0.64}$}, both consistent
with Gaussian initial conditions. We also constrain the
early-Universe energy injection amplitude to be
{$\alpha_{\rm inj} = -3.93^{+1.34}_{-0.99}$},
{with uncertainty reduced by a factor of $\sim\!2.6$ if
\textit{Planck} 2018 $\fnl$ constraint is applied as a prior.} Future surveys such as Simons Observatory and
{\it Euclid} will tighten these constraints further. Complementary
to conventional probes, this work provides the first ISW-tSZ
constraint on exotic energy injection and enables precision tests
of early-Universe physics while probing late-time gravitational
potential and thermal energy perturbations.
\end{abstract}
\maketitle
\fi

\section{Introduction}
 Cosmic inflation is widely regarded as the most compelling and physically plausible mechanism to generate the nearly scale-invariant primordial curvature fluctuations, which sourced the density perturbations to grow the cosmic structures that we observe today~\cite{Guth1981,Linde1982}. The basic inflationary model is the single-field slow-roll inflation model (SFSR), which predicts the Gaussian-distributed, adiabatic, almost scale-invariant primordial curvature fluctuations that any deviation from Gaussianity is small ($\mathcal{O}(10^{-3})$)~\cite{Liddle94,Langbein96,Forconi21,Auclair22,Lambiase23,Afshar23,Pozo24}. Departure from Gaussian initial conditions is conventionally described by the nonlinearity parameter $f_{\rm NL}$~\cite{Komatsu11,Planck2015_nG,Planck2018_nG}, which quantifies the quadratic correction to the Bardeen curvature perturbation $\Phi$ ($\Phi_{\rm H}$ in Bardeen~\cite{Bardeen1980}):
\begin{eqnarray}
\Phi(\mathbf{x}) = \Phi_{\rm L}(\mathbf{x}) + f_{\rm NL}\left[\Phi_{\rm L}^{2}(\mathbf{x}) - \langle \Phi_{\rm L}^{2}(\mathbf{x}) \rangle\right],
\end{eqnarray}
where $\Phi_{\rm L}$ denotes the Gaussian linear fluctuation with zero mean. This parametrization corresponds to the local shape of primordial non-Gaussianity, in which long- and short-wavelength modes are coupled in real space. Throughout this work, we therefore identify $f_{\rm NL} \equiv f_{\rm NL}^{\rm local}$ and use $f_{\rm NL}$ for brevity.

The quest to constrain PNG has advanced through complementary probes of the cosmic microwave background (CMB) and large-scale structures (LSS). The canonical CMB bispectrum analysis from \textit{Planck} yields the tightest constraint on the local PNG parameter, $f_{\rm NL} = -0.9 \pm 5.1$ (68\% confidence level (C.L.)), effectively limited by cosmic variance on large angular scales~\cite{Planck2018_parameters,Planck2018_nG}. In parallel, LSS surveys have emerged as alternative methods to cross-check the CMB constraints, constraining $f_{\rm NL}$ through the scale-dependent bias of galaxies at large scales~\cite{Dalal08,Ma2013,Li2017}. Recent analyses of the BOSS and eBOSS DR14 quasar samples constrain $f_{\rm NL} = -15^{+18}_{-21}$ (95\% C.L.) using redshift-weighting techniques~\cite{Castorina19}. The result is further refined in the final eBOSS DR16 analysis, reporting $f_{\rm NL} = -12 \pm 21$ (68\% C.L.) after systematic mitigation via neural-network-based cleaning~\cite{Rezaie21}. Using the bispectrum of BOSS galaxies, Ref.~\cite{Guido22} obtained $f_{\rm NL} = 7 \pm 31$ with an effective field theory approach. Joint power spectrum and bispectrum analyses provide complementary constraints that break degeneracies between galaxy bias and foreground systematics, yielding $f_{\rm NL} = -33 \pm 28$~\cite{Cabass22}. These LSS-derived $f_{\rm NL}$ values, while consistent with \textit{Planck} constraints, exhibit broader uncertainties that reflect the challenges of modeling nonlinear structure formation and redshift-space distortions.

The imprint of primordial non-Gaussianity extends beyond conventional
probes like the bispectrum or scale-dependent bias, manifesting in
correlations of different observables between vastly separated scales.
In the ultra-squeezed limit, local PNG (\(f_{\rm NL}^{\rm local}\))
modulates the dissipation of small-scale acoustic perturbations
(\(k \simeq 10^{2}-10^{4}\,{\rm Mpc}^{-1}\)) through their couplings
to the large-scale modes
(\(k \simeq 10^{-3}-10^{-2}\,{\rm Mpc}^{-1}\))~\cite{Pajer2012,Ganc12}.
This mechanism generates a unique correlation between
Compton-\(y\) distortions -- produced by energy injection at scales
\(1 \leq k \leq 50\,{\rm Mpc}^{-1}\) -- and large-angle CMB temperature
fluctuations~\cite{Chluba12,Emami2015}.

{%
Expanding Eq.~(1) about a long/short-wavelength split of the Gaussian
seed field and retaining terms linear in $f_{\rm NL}$, a long-wavelength
curvature mode $\Phi_{\rm L}$ rescales the local variance of the
small-scale modes by a factor $1+4 f_{\rm NL}\Phi_{\rm L}(\mathbf{x})$
(the standard squeezed-limit statement of the local ansatz; see
Ref.~\cite{DalalDore2008}). Silk damping of small-scale acoustic
modes at $1\,{\rm Mpc}^{-1}\lesssim k\lesssim 50~{\rm Mpc}^{-1}$ deposits energy
proportional to the small-scale power at those wavenumbers, so the
resulting Compton-$y$ distortion inherits the same modulation.
Cross-correlating this modulated $y$ with the Sachs-Wolfe temperature
anisotropy sourced by the same $\Phi_{\rm L}$ produces the
angular cross-power spectrum derived in Ref.~\cite{Chluba17}. The
resulting ISW-tSZ cross-correlation at $\ell\lesssim 200$ probes
$f_{\rm NL}^{\rm local}$ through the coupling between large-scale
potentials ($k_{\rm L}\lesssim 10^{-2}\,{\rm Mpc}^{-1}$) and the
small-scale acoustic modes whose dissipation generates the distortion,
with Fourier support at $1\lesssim k\lesssim 50\,{\rm Mpc}^{-1}$,
i.e.\ one to three decades above the CMB-bispectrum pivot
$k_\star\simeq 0.05\,{\rm Mpc}^{-1}$. This yields
the multipole expression}
\begin{eqnarray}
C^{y{\rm T}}_{\ell}({\rm PNG}) \simeq 12\, f_{\rm NL}\, \rho (\ell)
\left\langle y \right\rangle\,C^{\rm CMB}_{\ell},
\label{eq:cl_fNL}
\end{eqnarray}
{where the prefactor $12$ combines the
squeezed-limit mode-coupling amplitude with the Sachs-Wolfe transfer
of $\Phi_{\rm L}$ into $\Delta T/T$~\cite{Chluba17}. Because
$C^{y{\rm T}}_{\ell}({\rm PNG})\propto f_{\rm NL}\,\langle y\rangle$,
the signal vanishes for $f_{\rm NL}=0$: the mean primordial distortion
$\langle y\rangle$ by itself imprints no large-angle $y$--T
cross-correlation, so the measurement is a unique signature of $f_{\rm NL}$ observable that 
neither the $y$ auto- nor the $T$ auto-spectrum is.}
Here $C^{\rm CMB}_{\ell}$ is the dimensionless CMB temperature
power spectrum, and $\rho(\ell)\simeq 1.08 \times [1 - 2.2 \times
10^{-2}\,\ell - 1.72 \times 10^{-4}\, \ell^{2} +  2.0 \times 10^{-6}\,
\ell^3 - 4.56 \times 10^{-9}\, \ell^{4}]$ captures radiative transfer
effects relative to the Sachs-Wolfe plateau\footnote{This empirical fit
for the radiative transfer correction factor $\rho(\ell) =
C_\ell^{y{\rm T}}/C_\ell^{y{\rm T}, {\rm SW}}$ was derived by
\citet{Chluba17}. The function was obtained by fitting a polynomial to
the data for $C_\ell^{y{\rm T}}$ from \citet{Ganc12} (digitized from
their Figure 3) and normalizing by the analytic Sachs-Wolfe prediction.
The polynomial accurately captures the suppression of the signal due
to photon diffusion and the contributions from early and late-time
integrated Sachs-Wolfe effects over the multipole range $2 \leq \ell
\leq 200$.}. The functional form follows the squeezed-limit
prediction for local PNG~\cite{Emami2015,Chluba17}. In those works, $\langle y \rangle \simeq 4.2 \times 10^{-9}$ represents the \textit{primordial} Compton-$y$ distortion generated by acoustic damping of small-scale perturbations prior to recombination. This value arises from theoretical modeling of energy dissipation during the cosmic dark ages and reflects the expected amplitude under standard $\Lambda$CDM cosmology with Gaussian initial conditions. 

Recent advances in cross-correlation techniques enable a direct observational test of this mechanism. In Ref.~\cite{Ibitoye24}, we achieved a \( 3.6\sigma \) detection of the correlation between the ISW effect and tSZ signal, tracing the time evolution of gravitational potentials and electron pressure fluctuations, respectively. This detection suggests that, although the electron pressure is mostly associated with galaxy clusters on scales of $\sim 1\text{--}10\,{\rm Mpc}$, its integrated signal is correlated with the ISW effect on very large scales, up to $\sim 1\,h^{-1}\,{\rm Gpc}$~\cite{ISW_tSZ_011,Creque-Sarbinowski16}. Crucially, while accessing larger scales ($\ell \leq 200$) that the signature of PNG is most pronounced, this approach circumvents key limitations of galaxy surveys---such as redshift-dependent systematics~\cite{DES2023,Saraf2024} and nonlinear modeling uncertainties \cite{Guido22,Cabass22}--thus avoids the traditional challenge that large-scale structure (LSS) bispectrum analyses face. Because the ISW–tSZ correlation encodes PNG signatures through its sensitivity to coupled large- and small-scale perturbations, in this work we leverage the $y$T correlation function measured in Ref.~\cite{Ibitoye24} to derive constraints on $\fnl$. In doing so, we incorporate both fluctuation-dependent energy deposition and the mean $\langle y \rangle$ contribution, enabling a joint analysis of three effects: primordial non-Gaussianity, energy injection, and the large-scale intergalactic medium.

Our analysis adopts a spatially-flat $\Lambda$CDM cosmology model with primordial fluctuation parameters fixed to {\it Planck}-2018 best-fitting values (\( n_{\rm s}=0.965 \), \( \tau=0.0540 \), \( \ln(10^{10}A_{\rm s})=3.043 \)~\cite{Planck2018_parameters}). But we do vary the large-scale structure parameters ($\Omega_{\rm m}$, $h$, $\sigma_{8}$) alongside with $\fnl$ because they affect the prediction of the intergalactic medium {(IGM)} part. Our approach marks the first application of ISW-tSZ cross-correlations to primordial non-Gaussianity and early energy injection, offering a complementary pathway to advance cosmology beyond the systematic ceilings of current LSS and CMB bispectrum methods.\\

 \begin{figure*}
	\centering
        \includegraphics[width=7.3in]{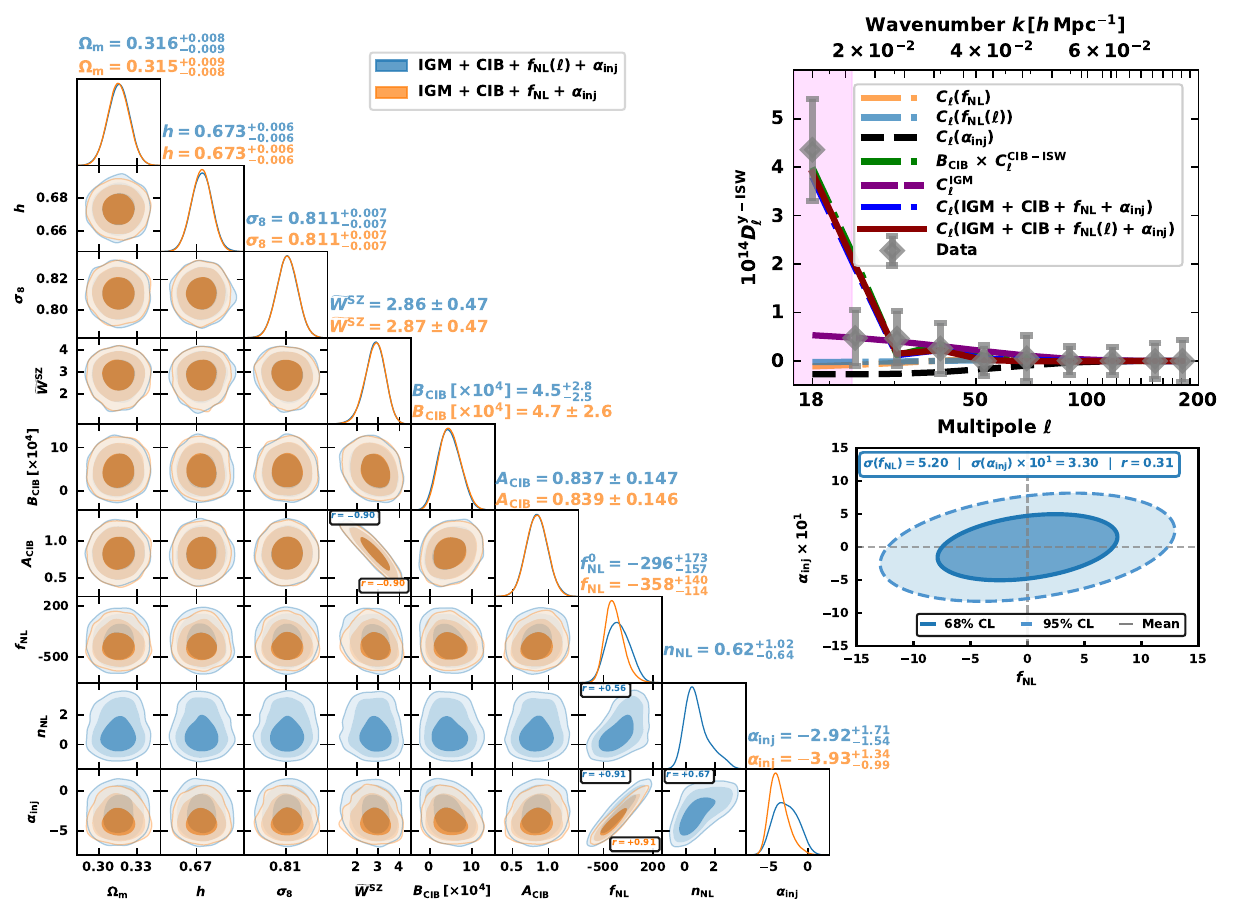}
\caption{{\it Lower-left triangle}: Posterior distributions from MCMC with the full covariance matrix for both scale-dependent {(dodger-blue)} and scale-independent scenarios {(orange)}, {using Gaussian priors on cosmological parameters ($\Omega_{\rm m}$, $h$, $\sigma_8$) from \textit{Planck} 2018 TT,TE,EE+lowE+lensing~\cite{Planck2018_parameters} (see also Table~\ref{tab:estimates_horizontal_no_S8})}. For the scale-independent PNG, $\fnl$ is being sampled as a constant parameter from Eq.~(\ref{eq:cl_fNL}) in the MCMC; while in the scale-dependent case, both $\fnl^{0}$ and $n_{\rm NL}$ (Eq.~(\ref{eq:fnl_scaled})) are sampled as free parameters. In the latter case, the sub-plot of $\fnl$ constraint indicates the constraint for $\fnl^{0}$. Pearson correlation coefficients $|r| \geq 0.5$ are annotated for both scenario to highlight strong parameter couplings. {\it Upper-right panel}: Different contributions of the ${y\rm T}$ cross-correlation power spectrum ($D_{\ell}\equiv \ell(\ell+1)C_{\ell}/2 \pi$) according to the best-fitting parameters in Table~\ref{tab:estimates_horizontal_no_S8}. The top boundary $x$-axis uses the transformation $k=(\ell +0.5)/\chi(z)$ to show the comoving wavenumber ($\chi\simeq 2.7\times 10^{3}\,h^{-1}{\rm Mpc}$). The purple curve captures the IGM ($C^{\rm tSZ-ISW}_{\ell}$ in Eq.~(\ref{eq:cl-ISW_y})), while the green line represents the CIB contribution weighted by the $B_{\rm CIB}$ parameter (Eq.~(\ref{eq:cl_ISW-tSZ})). The orange and dodger-blue curves are the scale-independent and scale-dependent cases of PNG contribution (Eqs.~(\ref{eq:cl_fNL}) and (\ref{eq:fnl_scaled})). The black line represents the average energy injection from the early Universe (Eq.~(\ref{eq:cl_E_injection})). Because the first data point ($\ell_{\rm eff}=18$) is heavily dominated by CIB, we did a test of removing this point in our MCMC analysis and the results are discussed in Section 2. {\it Lower-right panel}: Projected 2D constraint on $\fnl$--$\alpha_{\rm inj}$ parameter space with SO and {\it Euclid} surveys, calculated via Fisher analysis (Eq.~(\ref{eq:Fij})).}
\label{fig:y_ISW_fnl constraints}
\end{figure*}

\section{Modeling \& Data Analysis}
\label{sec:modeling}
We now model the {\it total} contribution to Compton-$y$ distortion and ISW cross-correlation power spectrum. The first contribution is what is shown in Eq.~(\ref{eq:cl_fNL}), the energy injection due to the dissipation of acoustic modes on scales of $1 \leq k \leq 50\,{\rm Mpc}^{-1}$, because of the existence of the squeezed-limit PNG. We model $\fnl$ both as a single parameter and as a function of angular scales, i.e.
\begin{eqnarray}
f_{\rm NL}(\ell) = f^{0}_{\rm NL} \bigg(\frac{\ell}{\ell_{0}}\bigg)^{n_{\rm NL}},
\label{eq:fnl_scaled}
\end{eqnarray} 
where $f^{0}_{\rm NL}$ is the normalization at the pivot scale $\ell_{0}=200$, and $n_{\rm NL}$ is the power-law index. We refer to these as the scale-independent and scale-dependent models, respectively.

The second contribution is the large cosmic scale, IGM component, {which arises due to} the correlation between the underlying gravitational potential field (ISW) and the warm-hot IGM. For this part, the Limber approximation is sufficient to model the data, because it is accurate on linear regimes where our data are binned\footnote{This approximation assumes that the line-of-sight integral over $k_{\parallel}$ is only effective for low-$k_{\parallel}$, due to the cancellation of rapid varying of high-$k_{\parallel}$ modes. This is easy to satisfy for high-$\ell$ modes because $P(\sqrt{k^{2}_{\parallel}+(\ell/\chi)^{2}})\simeq P(\ell/\chi)$, which otherwise is not accurate for low-$\ell$ modes~\cite{Dodelson2020}.}. This contribution is modelled with two parts~\cite{Ibitoye24}
\begin{eqnarray}
  \label{eq:cl_ISW-tSZ}
  C^{y{\rm T}}_{\ell}({\rm ISW}) = C^{\rm tSZ-ISW}_{\ell} + B_{\rm CIB}\,C^{\rm CIB_{avg}-ISW}_{\ell}\,,
\end{eqnarray}
where the second term accounts for residual cosmic infrared background (CIB) contamination via the CIB-ISW cross correlation {weighted by the dimensionless amplitude parameter $B_{\rm CIB}$, which is marginalized over in our analysis.} We adopt the averaged CIB-ISW cross power spectrum in Sec. 4.4 in~\citet{Ibitoye24} {(see Fig.~7 therein)} and {we} vary $B_{\rm CIB}$ (in unit of ${\rm sr}\cdot {\rm Myr}^{-1}$) in our likelihood chain. The first term is the principle term, which is theoretical ISW-tSZ cross-power spectrum, for which we can utilize the general formula for auto- and cross-correlations of $X$ and $Y$ variables 
\begin{eqnarray}
 \label{eq:cl-ISW_y}
C^{XY}_{\ell} &=&  \int {\rm d}z \,\Delta^{X}_{\ell}(z) \Delta^{Y}_{\ell}(z) P_{\rm m}\left(k\equiv \frac{\ell+1/2}{\chi},z\right),
\end{eqnarray} 
where $X$ and $Y$ can be both ISW for the auto-power spectrum ($C^{\rm ISW}_{\ell}$), or both tSZ for $C^{\rm tSZ}_{\ell}$, or one ISW and one tSZ for the cross-power spectrum ($C^{\rm tSZ-ISW}_{\ell}$). ${\rm d}\chi=c\,{\rm d}z/H(z)$ is the differential comoving distance, and $P_{\rm m}\left(k,z\right)$ is the linear matter power spectrum at redshift $z$. The kernels are 
\begin{eqnarray}
\Delta^{\rm ISW}_{\ell}(z)&=& \frac{3\Omega_{\rm m}H^{2}_{0}}{c^{2}\left(\ell +1/2 \right)^{2}} \frac{\chi(z)}{D(z)}\frac{\rm d}{{\rm d}z} \left(\frac{D(z)}{a(z)} \right) \,\nonumber \\
\Delta^{\rm SZ}(z) &=& \left(\frac{k_{\rm B}\sigma_{\rm T}}{m_{\rm e}c^{2}} \right) \frac{\widetilde{W}^{\rm SZ}}{\chi}\;,
\label{eq:tilde-delta}
\end{eqnarray} 
where $D(z)$ is the linear growth factor and
\begin{eqnarray}
\widetilde{W}^{\rm SZ} &=&   b_{\rm gas} \left(\frac{\bar{n}_{\rm e 0}}{1\,{\rm m}^{-3}} \right)\left(\frac{k_{\rm B}T_{\rm e}(0)}{0.1\,{\rm keV}} \right),
\end{eqnarray} 
in which $b_{\rm gas}$ is the gas bias at redshift $0$.

The third contribution comes from the average energy injections of the early Universe. While the Doppler and potential driving are injections on large scales, the perturbed thermalization and anisotropic heating are relevant on small scales. As shown in {Fig}.~27 of Ref.~\cite{Kite23}, the effect of energy injections to $\Theta \times y$ power spectrum ($\Theta\equiv \Delta T/T$ is the temperature variation) at different injection times are calculated with a fixed energy release $\Delta \rho/\rho = 3\times 10^{-5}$. For simplicity, we adopt the $\mathrm{tSZ}\Theta$ spectrum (orange curve) from {Fig}.~27 of \citet{Kite23}, hereafter denoted $C^{\rm inj}_{\ell}$, as a template model for the average energy injection spectrum. {We multiply it by a free parameter $\alpha_{\rm inj}$, to rescale its overall amplitude in the contribution to} $C^{y{\rm T}}_{\ell}$ and vary this parameter in the subsequent likelihood analysis\footnote{In practice, we also consider the alternative energy injection scenarios by adopting different blue curves in {Fig}.~27 of Ref.~\cite{Kite23} but the resultant {constraint} does not change much; therefore, we present here the ${\rm tSZ}\Theta$ case as {a} representative example.}:
\begin{eqnarray}
C^{y{\rm T}}_{\ell} ({\rm Inj})&=& \alpha_{\rm inj}\,   C^{\rm inj}_{\ell} .
\label{eq:cl_E_injection}
\end{eqnarray}

Combining three parts of contribution (Eqs.~(\ref{eq:cl_fNL}), (\ref{eq:cl-ISW_y}) and (\ref{eq:cl_E_injection})), our theoretical model is
\begin{eqnarray}
  \label{eq:cl_ISW-tSZ_total}
  C^{y{\rm T}}_{\ell} = C^{y{\rm T}}_{\ell}({\rm ISW})+ C^{y{\rm T}}_{\ell}({\rm PNG}) + C^{y{\rm T}}_{\ell}({\rm Inj}),
\end{eqnarray}
which in total contains eight free parameters ($\vec{\theta}=\Omega_{\rm m}, h, \sigma_8,\widetilde{W}^{\rm SZ}, B_{\rm CIB}, A_{\rm CIB}, f_{\rm NL}, \alpha_{\rm inj}$), where the $f_{\rm NL}$ can be unfolded into another 2D parameter space (Eq.~(\ref{eq:fnl_scaled})). {For our fiducial analysis, we adopt Gaussian priors on the cosmological parameters based on \textit{Planck} 2018 TT,TE,EE+lowE+lensing constraints: $\Omega_{\rm m} = 0.3153 \pm 0.0073$, $h = 0.6736 \pm 0.0054$, and $\sigma_8 = 0.8111 \pm 0.0060$~\cite{Planck2018_parameters}.}
{The first three parameters ($\Omega_{\rm m}, h, \sigma_8$) set up the cosmological background and perturbations, while $\widetilde{W}^{\rm SZ}$ sets the amplitude of the gaseous signal (Compton-$y$ and therefore $C_{\ell}^{\rm tSZ-ISW}$). Parameters $B_{\rm CIB}$ and $A_{\rm CIB}$ characterize residual CIB contamination in the measured $C_\ell^{y{\rm T}}$ and $C_\ell^{yy}$ spectra, because $C^{y{\rm T}}_{\ell}({\rm ISW})$ is modeled via Eq.~(\ref{eq:cl_ISW-tSZ}) and $C^{yy}_{\ell} = (C^{\rm tSZ}_{\ell}+A_{\rm CIB}C^{\rm CIB}_{\ell})$. Here $C^{\rm tSZ}_{\ell}$ is calculated via Eq.~(\ref{eq:cl-ISW_y}) with tSZ kernel and $C^{\rm CIB}_{\ell}$ is the CIB power spectrum~\cite{Ibitoye24}.} The seventh parameter is $\fnl$, for which we consider for both scale-independent case and scale-dependent case (Eq.~(\ref{eq:fnl_scaled})) therefore the latter case introduces an additional parameter ($n_{\rm NL}$). The last parameter $\alpha_{\rm inj}$ represents the total average amplitude of perturbed thermalization and anisotropic heating. Our goal is to place a novel constraint on PNG and the amplitude of the average energy injection from the early Universe, but we vary the other large-scale structure parameters and CIB parameters to assess the degeneracies among the parameters. Therefore for full constraints, we vary all physical and nuisance parameters simultaneously in the likelihood while performing the joint fitting to $C_\ell^{\rm TT}, C_\ell^{y{\rm T}}$ and $C_\ell^{yy}$. For conditional constraints, we fix other parameters to their fiducial values but only varying $\fnl$ parameters and $\alpha_{\rm inj}$ of interest (Table~\ref{tab:estimates_horizontal_no_S8}).

\begin{table*}
\centering
\renewcommand{\arraystretch}{1.3}
\caption{Constraints of the parameters in the model
($C^{y{\rm T}}_{\ell}$ in Eq.~(\ref{eq:cl_ISW-tSZ_total})) which contains
PNG, IGM and energy injection contributions, where the quoted error is
$68\%$ C.L. unless otherwise noted. \textbf{Full constraints} -- all
listed parameters are free to run and sampled in the likelihood chain.
{The upper block ({\it Planck} cosmology priors) comprises the rows $f_{\rm NL}$~(SI),
$f_{\rm NL}(\ell)$~(SD) and $\alpha_{\rm inj}\!\ge\!0$~(SI), imposes Gaussian
priors on the cosmological parameters from \textit{Planck} 2018
TT,TE,EE+lowE+lensing~\cite{Planck2018_parameters}, as listed in the
first ``Prior'' row.} {The row labelled
$\alpha_{\rm inj}\!\ge\!0$~(SI) repeats the fiducial scale-independent
chain (row~1) with the flat prior
$\alpha_{\rm inj}\sim\mathcal{U}[0,5]$, all other priors unchanged (see Appendix~\ref{app:alpha_positive}.)} 
{The lower block, comprising the row
$\mathcal{P}(f_{\rm NL})$, retains the fiducial flat cosmology priors (listed in the second ``Prior'' row), and adds the Gaussian prior
$\mathcal{P}(f_{\rm NL})=\mathcal{N}(-0.9,\,5.1^{2})$ from
\textit{Planck} 2018 CMB bispectrum~\cite{Planck2018_nG}, while all
other parameters remain flat.} {Additional chains
that further marginalise over Galactic dust and synchrotron leakage
amplitudes ($\mathcal{P}(f_{\rm NL})+\text{d/s}$~(SI) and (SD)) are
reported separately in Table~\ref{tab:foregrounds_appendix} of
Appendix~\ref{sec:foregrounds}.} \textbf{Conditional constraints}:
only the PNG and energy injection parameters are sampled, i.e.\
$f_{\rm NL}$, $n_{\rm NL}$, $\alpha_{\rm inj}$ and $B_{\rm CIB}$, while
the other cosmological and foreground parameters are fixed to the
fiducial values ($\Omega_{\rm m}=0.3157$, $h=0.6731$,
$\sigma_{8}=0.8151$ according to
Ref.~\cite{Planck2018_parameters}, and $\widetilde{W}^{\rm SZ}=2.89$
and $A_{\rm CIB}=0.838$). Priors are uniform over the indicated ranges
except the Gaussian prior $\mathcal{P}({\fnl})$.
{$^{\ast}$~One-sided $95\%$ upper limit from the
$\alpha_{\rm inj}\!\ge\!0$ posterior.}}
\label{tab:estimates_horizontal_no_S8}

\scaledtable[1.0\textwidth]{%
\begin{tabular}{l|ccccccccc}
\hline\hline
$\vec{\theta}$& $\Omega_{\rm m}$ & $h$ & $\sigma_8$ & $\widetilde{W}^{\rm SZ}$ & $f_{\rm NL}$ & $n_{\rm NL}$ & $\alpha_{\rm inj}$ & $B_{\rm CIB}\,[\times 10^{4}]$ & $A_{\rm CIB}$ \\
\hline
\multicolumn{10}{c}{\textbf{Full Constraints}}\\
\hline
Prior & {$\mathcal{N}(0.3153,0.0073^2)$} & {$\mathcal{N}(0.6736,0.0054^2)$} & {$\mathcal{N}(0.8111,0.0060^2)$} & $\mathcal{U}[0.02,10]$ & $\mathcal{U}[-5000,5000]$ & $\mathcal{U}[-3,3]$ & $\mathcal{U}[-5,5]$ & $\mathcal{U}[-5,5]$ & $\mathcal{U}[0,5]$ \\
$f_{\rm NL}$~(SI) & {$0.3152^{+0.009}_{-0.008}$} & {$0.6732 \pm 0.0060$} & {$0.8108 \pm 0.0070$} & {$2.87 \pm 0.47$} & {$-358^{+140}_{-114}$} & $-$ & {$-3.93^{+1.34}_{-0.99}$} & {$4.7\pm2.6$} & {$0.84 \pm 0.146$} \\
$f_{\rm NL}(\ell)$~(SD)   & {$0.3160 ^{+0.008}_{-0.009}$} & {$0.6730 \pm 0.0060$} & {$0.8109 \pm 0.0070$} & {$2.86 \pm 0.47$} & {$-296^{+173}_{-157}$} & {$0.62^{+1.02}_{-0.64}$} & {$-2.92^{+1.71}_{-1.54}$} & {$4.5^{+2.8}_{-2.5}$} & {$0.84 \pm 0.147$} \\
{$\alpha_{\rm inj}\!\ge\!0$ (SI)}
    & {$0.3157 \pm 0.0074$}
    & {$0.6724 \pm 0.0054$}
    & {$0.8108 \pm 0.0060$}
    & {$2.60 \pm 0.47$}
    & {$86 \pm 73$}
    & {$-$}
    & {$<1.79^{\ast}$}
    & {$2.9 \pm 1.8$}
    & {$0.91 \pm 0.13$}\\
\hline
Prior & {$\mathcal{U}[0.20,0.42]$} & {$\mathcal{U}[0.60,0.80]$} & {$\mathcal{U}[0.64,0.90]$} & $\mathcal{U}[0.02,10]$ & ${\mathcal{N}(-0.9,\,5.1^2)}$ & $\mathcal{U}[-3,3]$ & $\mathcal{U}[-5,5]$ & $\mathcal{U}[-5,5]$ & $\mathcal{U}[0,5]$ \\
$\mathcal{P}(f_{\rm NL})$ & $0.317^{+0.074}_{-0.074}$ & $0.635^{+0.044}_{-0.032}$ & $0.745^{+0.099}_{-0.084}$ & $2.77^{+0.66}_{-0.63}$ & $-1.11^{+6.18}_{-6.13}$ & $-$ & $-0.50^{+0.54}_{-0.54}$ & $3.9^{+2.74}_{-2.48}$ & $0.868^{+0.145}_{-0.149}$ \\
\hline
\multicolumn{10}{c}{\textbf{Conditional Constraints}}\\
\hline
$f_{\rm NL}$ & \multicolumn{1}{c}{---} & \multicolumn{1}{c}{---} & \multicolumn{1}{c}{---} & \multicolumn{1}{c}{---} & $-402^{+145}_{-116}$ & $-$ & $-3.90^{+1.33}_{-0.98}$ & $1.1 \pm 0.36$ & \multicolumn{1}{c}{---} \\
$f_{\rm NL}(\ell)$   & \multicolumn{1}{c}{---} & \multicolumn{1}{c}{---} & \multicolumn{1}{c}{---} & \multicolumn{1}{c}{---} & $-302^{+191}_{-181}$ & $0.58^{+1.61}_{-1.07}$ & $-2.61^{+1.95}_{-1.82}$ & $1.0 \pm 0.37$ & \multicolumn{1}{c}{---} \\
\hline
\end{tabular}}
\end{table*}
We construct the {\it data} vector from the measured $C^{y{\rm T}}_{\ell}$, $C^{\rm TT}_{\ell}$ and $C^{yy}_{\ell}$ as $\mathbf{C}^{\rm obs}_{\ell}\equiv (C^{\rm TT}_{\ell}, C^{y{\rm T}}_{\ell}, C^{yy}_{\ell})$, and use their covariance matrices from Ref.~\cite{Ibitoye24} to calculate the $\chi^{2}$ and the subsequent likelihood function\footnote{See Table~1 in Ref.~\cite{Ibitoye24} for the data, and section 2 for a detailed description of the data products and covariance matrices)}. For $\mathbf{C}^{\rm th}_{\ell}$, we use the theoretical ISW auto-power spectrum ($C^{\rm ISW}_{\ell}$ from Eq.~(\ref{eq:cl-ISW_y}) in which kernel $\Delta^{\rm ISW}_{\ell}$ from Eq.~(\ref{eq:tilde-delta})), $C^{y{\rm T}}_{\ell}$ from Eq.~(\ref{eq:cl_ISW-tSZ_total}) and $C^{yy}_{\ell}$ mentioned above. The $\chi^{2}$ function then becomes $-2\ln\mathcal{L}=\chi^{2}(\vec{\theta})=\sum_{\ell \ell'}(\mathbf{C}^{\rm obs}_{\ell}-\mathbf{C}^{\rm th}_{\ell}(\vec{\theta}))\left({\rm Cov}^{\rm {Tot}} \right)^{-1}_{\ell \ell'} (\mathbf{C}^{\rm obs}_{\ell'}-\mathbf{C}^{\rm th}_{\ell'}(\vec{\theta}))$, which is being minimized in our likelihood chain. The measured binned power spectra have 10 band powers, ranging in $\ell \in [16, 190]$. To increase the statistical weight on the multipoles and test if {this change results} in an appreciable impact on the analysis, we increased the number of bins in the whole range of $\ell$ to 44 using the \texttt{NaMaster} binning scheme through the \textsc{NmtBins} subroutine, which allows to control $\ell$-weighting with custom-made band powers. We recomputed the power spectra in this way and repeated the Markov-Chain Monte Carlo (MCMC) method to estimate the parameters. {The resultant constraints on $f_{\rm NL}$ and $\alpha_{\rm inj}$ shift by less than 0.3$\sigma$ compared to the fiducial 10-bin analysis, and the best-fit $\chi^2$ changes by $\Delta\chi^2 < 0.5$, indicating no significant impact.} In addition, because the first bin ($\ell_{\rm eff} = 18.0$) is dominated by CIB contamination, we performed a test excluding it and using only the 9 higher-$\ell$ bins ($\ell_{\rm eff} \in [21,190]$). {The inferred parameters shift by at most 0.4$\sigma$, and the goodness-of-fit remains consistent ($\Delta\chi^2 \simeq 0.7$), confirming that our results are robust to the treatment of the lowest-$\ell$ point.} { Because the lowest-$\ell$ bin ($\ell = 16$–$25$; pink band in Fig.~\ref{fig:y_ISW_fnl constraints}) contributes only $6.5\%$ of the total information, and any bias from the Limber approximation is expected to be well below $3\%$ (see the Fisher forecast in Sec.~\ref{sec:summary}), we consider the Limber approximation to be sufficiently accurate and not expected to appreciably affect the resulting parameter constraints.}

\section{Results and Discussion}
\label{sec:results}
The full constraints are summarized in Fig.~\ref{fig:y_ISW_fnl constraints} and Table~\ref{tab:estimates_horizontal_no_S8}. The lower-left triangle of Fig.~\ref{fig:y_ISW_fnl constraints} displays the 2D and 1D marginalized posterior distributions from our MCMC analysis, in which the cosmological parameters (\(\Omega_{\rm m}\), \(h\), \(\sigma_8\)) are {recovered at their input \textit{Planck} 2018 values, as expected given the Gaussian priors imposed on them; the ISW-tSZ bandpowers do not tighten these directions appreciably}. The upper-right panel of Fig.~\ref{fig:y_ISW_fnl constraints} decomposes the best-fit ISW-tSZ cross-power spectrum into different physical components. The dominant contribution comes from the intergalactic medium (IGM; purple curve), which accounts for approximately 90\% of the total signal and is responsible for the 3.6$\sigma$ detection reported in \citet{Ibitoye24}. The CIB term (green line) is subdominant but non-negligible at low multipoles (\(\ell < 40\)). In contrast, both the PNG contributions (orange for scale-independent, dodger-blue for scale-dependent) and the average energy injection term (black line) are marginal, consistent with no detection. Quantitatively, we  obtain the following 68\% confidence-level constraints from the full parameter fit (Table~\ref{tab:estimates_horizontal_no_S8}):  
\begin{eqnarray}
{f_{\rm NL} = -358^{+140}_{-114}} \quad \text{(scale-independent)} \nonumber \\
\left.
\begin{aligned}
 {f^0_{\rm NL}} &= {-296^{+173}_{-157}}  \\
    {n_{\rm NL}} &= {0.62^{+1.02}_{-0.64}}
  \end{aligned}
  \right\} \quad \text{(scale-dependent).}
\end{eqnarray}

{%
We emphasise that $f_{\rm NL}$ is not a positive-definite parameter, so
the sign of the central value is itself meaningful. A notable feature
of the fiducial scale-independent fit is that both $f_{\rm NL}$ and
$\alpha_{\rm inj}$ sit on the negative side of zero, at $2.6\sigma$ and
$2.9\sigma$ respectively, quoting in each case the error bar on the
side facing zero. Imposing the physical prior
$\alpha_{\rm inj}\!\ge\!0$ (third row of
Table~\ref{tab:estimates_horizontal_no_S8}) returns
$\alpha_{\rm inj}<1.79$ ($95\%$ C.L.) and $f_{\rm NL}=86\pm 73$,
i.e.\ both parameters recover consistency with zero at $1.2\sigma$;
the same happens when the \textit{Planck} bispectrum prior is imposed
instead (row $\mathcal{P}(f_{\rm NL})$). The negative central values
are therefore a symptom of the $f_{\rm NL}$--$\alpha_{\rm inj}$
degeneracy rather than a detection of exotic physics; we demonstrate
this explicitly in Appendix~\ref{app:alpha_positive}, verify in
Appendix~\ref{app:IGM_degeneracy} that $\widetilde{W}^{\rm SZ}$ is not
absorbing either signal, and show in Appendix~\ref{sec:foregrounds}
that the conclusion survives Galactic foreground marginalisation.} {{Turning to scale dependence, a $\Delta\chi^2$ test shows all values in $n_{\rm NL} \in [-3,\,2.5]$ yield $\Delta\chi^2<1$, which indicates the insufficient leverage of our current measurement to constrain the running parameter. Nevertheless, $\fnl$ scale-dependency is theoretically well motivate from multi-field inflation \cite{Byrnes10}, features in the inflationary potential \cite{Chen07}, and non-Bunch-Davies vacua \cite{Holman08}, so it is worthwhile to pursue with future probes.}}

{We assess the goodness-of-fit of our models using the full bandpower covariance derived from $900$ Gaussian simulations ($N_{\rm sim}/N_{\rm bins}=90$).
The scatter between bandpowers is smaller than the individual error bars because adjacent bandpowers are positively correlated through the ISW filter and beam window; the covariance matrix fully accounts for this. Degrees of freedom count only parameters constrained by the $y$--$T$ bandpowers themselves; nuisance amplitudes constrained by the $C_\ell^{yy}$ or $C_\ell^{\rm TT}$ channels of the joint fit are treated as effectively fixed. Under this convention the IGM-only model gives $\chi^2/\mathrm{dof}=18.0/6$ (Probability of Exceed (PTE)~$=0.006$), the CIB-only template regression gives $8.09/9$ (PTE~$=0.53$), the joint CIB$+$dust$+$sync template regression gives $4.49/7$ (PTE~$=0.72$), and the full MCMC best-fit (IGM$+$CIB$+f_{\rm NL}+\alpha_{\rm inj}$) gives $7.60/3$ (PTE~$=0.055$); a PTE of $0.055$ corresponds to a $1.9\sigma$ deviation, which is within the expected range for a 10-bandpower fit with seven free parameters and does not indicate model failure; the full breakdown is collected in Table~\ref{tab:gof}.}

For the energy injection amplitude $\alpha_{\rm inj}$, we introduce a scaling parameter multiplying the template of Eq.~(\ref{eq:cl_E_injection}). {Positive values correspond to additional heating processes such as particle decay or primordial magnetic fields, while negative values correspond to a suppression of energy deposition relative to the fiducial template rather than to an unphysical absolute energy}.
Our measurement yields
{$\alpha_{\rm inj} = -3.93^{+1.34}_{-0.99}$} ($68\%$~C.L.),
{i.e.\ $2.9\sigma$ below zero in the fiducial chain; as shown above this
excursion disappears once the physical prior
$\alpha_{\rm inj}\!\ge\!0$ is imposed, and the resulting one-sided
bound $\alpha_{\rm inj}<1.79$ ($95\%$ C.L.) is the statement we regard as physical}. We further impose a \textit{Planck} bispectrum prior on $f_{\rm NL}$, modeled as a Gaussian distribution with mean $-0.9$ and standard deviation $5.1$, which results in a reduction of the uncertainty on $\alpha_{\rm inj}$ by a factor of {$2.5$} without significantly affecting the other parameters (Table~\ref{tab:estimates_horizontal_no_S8}). This demonstrates that external probes of PNG (e.g., CMB or galaxy bispectra) can be leveraged to constrain $\alpha_{\rm inj}$ within this framework.

{The constraint derived here on $f_{\rm NL}^{\rm local}$ probes the
squeezed-limit mode coupling in a wavenumber window fundamentally
distinct from that of the two other observational routes to
primordial non-Gaussianity. The {\it Planck}-2018 CMB
bispectrum~\cite{Planck2018_nG} is anchored at the CMB pivot
$k_{\star}\simeq 0.05~{\rm Mpc}^{-1}$, and current galaxy
scale-dependent-bias analyses~\cite{Castorina19,Cabass22}
probe $k\lesssim 0.1~{\rm Mpc}^{-1}$. In contrast, the ISW-tSZ
cross-correlation at $\ell\le 200$ inherits its
$f_{\rm NL}$-sensitivity through the small-scale acoustic-dissipation
window that sources the primordial Compton-$y$ distortion, i.e.\ from
modes with $1\lesssim k\lesssim 50~{\rm Mpc}^{-1}$
\cite{Chluba12,Emami2015,Chluba17}. Weighting that window
logarithmically gives an effective pivot
$k_{y}\equiv(k_{\rm min}k_{\rm max})^{1/2}\simeq 7~{\rm Mpc}^{-1}$,
which we use in what follows purely as a compact label for the
dissipation window rather than as a precisely determined scale. Our
sensitivity therefore lies between one and three decades above
$k_{\star}$, with $k_{y}/k_{\star}\simeq 1.4\times 10^{2}$. While the
numerical error bar reported here is broader than the \textit{Planck}
bispectrum's, we are not aware of any published constraint on
$f_{\rm NL}^{\rm local}$ at $k\gtrsim 1~{\rm Mpc}^{-1}$; the present
result is therefore the first at these wavenumbers from any two-point
CMB observable, and it is complementary to, rather than in competition
with, the CMB-bispectrum and galaxy-bias results.}

To further isolate the constraining power of ISW-tSZ cross-correlation to the PNG and $\alpha_{\rm inj}$, we did a conditional likelihood fitting, by fixing the cosmological parameter values to the {\it Planck}-2018 best-fitting value and the $\widetilde{W}^{\rm SZ}$ and $A_{\rm CIB}$ to be the best-fitting full constraints value (Table~\ref{tab:estimates_horizontal_no_S8}). {The conditional constraints, $f_{\rm NL}=-402^{+145}_{-116}$ and
$\alpha_{\rm inj}=-3.90^{+1.33}_{-0.98}$, reproduce the same negative
excursion as the full fit and at comparable significance, showing that
it originates in the $y{\rm T}$ likelihood itself rather than in the
marginalisation over the cosmological and CIB parameters; as in the
full fit, it is removed once the physical prior
$\alpha_{\rm inj}\!\ge\!0$ is imposed
(Appendix~\ref{app:alpha_positive}). The physical interpretation is
therefore unchanged.}
One thing to notice is that strong degeneracies between astrophysical and foreground parameters: \(\widetilde{W}^{\rm SZ}\) and \(B_{\rm CIB}\) are anti-correlated (\(r \simeq -0.58\)), reflecting the trade-off between the IGM signal and residual CIB contamination. Most notably, the PNG and energy injection parameters exhibit tight couplings: \(f_{\rm NL}\) and \(n_{\rm NL}\) are positively correlated (\(r \simeq +0.92\)), as well as \(n_{\rm NL}\) and \(\alpha_{\rm inj}\) (\(r \simeq +0.93\)), indicating that current data may not give individual constraints on scale-dependent primordial non-Gaussianity and anomalous energy injection. {A broader multipole coverage from future surveys is required to break the parameter degeneracy.}

{Nonetheless, our measurement of $\alpha_{\rm inj}$ places meaningful limits on exotic energy injection in the early Universe ($z \gtrsim 10^{3}$), beyond the existing limits from COBE/FIRAS. Such energy injection may arise from exotic processes such as decaying dark matter or primordial black holes~\cite{Chluba15}, whose physical nature remains to be probed. Our analysis therefore provides the first observational constraint that simultaneously probes late-time ionization and thermal pressure perturbations with early-Universe physics through the ISW–tSZ cross-correlation.}

\section{Summary and Outlook}
\label{sec:summary}
In this paper, we present a novel approach to constrain PNG and the energy injection in the early Universe {by using the ISW effect and tSZ effect cross-correlation measurement}~\cite{Ibitoye24}. Our analysis accounts for contributions from the IGM, the CIB, PNG, and early-Universe energy injection to the ISW--Compton-$y$ correlation. We adopt both scale-independent and scale-dependent models of local PNG, using MCMC to fit the parameters. We obtain 68\% C.L. constraints: 
{$f_{\rm NL} = -358^{+140}_{-114}$} (scale-independent), 
and {$f^0_{\rm NL} = -296^{+173}_{-157}$}, {$n_{\rm NL} =0.62^{+1.02}_{-0.64}$} (scale-dependent). A significant detection of $f^0_{\rm NL}$ at large amplitude would rule out the broad class of single-field, slow-roll inflationary models with canonical kinetic terms and Bunch-Davies initial conditions~\cite{KomatsuE02}, pointing instead to multi-field or non-standard dynamics. {Because there is a parameter degeneracy between $f_{\rm NL}$ and $\alpha_{\rm inj}$ (Appendix~\ref{app:alpha_positive}), we do not use their sign to
discriminate between inflationary models: our constraints remain
compatible both with scenarios predicting large {positive}
non-Gaussianity, such as G-inflation~\cite{ZhangF21} or rapid-turn
trajectories~\cite{Iarygina24}, which typically yield
$f_{\rm NL}\gtrsim\mathcal{O}(10)$ under standard templates, and with
those predicting suppressed or negative PNG, such as two-field warm
inflation~\cite{WangY19} or Higgs inflation~\cite{Rubio18}.}
The lack of significant scale dependence further suggests that deviations from scale invariance in PNG are small, narrowing the parameter space for multi-field couplings and potential features. For the energy injection amplitude we obtain
{$\alpha_{\rm inj} =-3.93^{+1.34}_{-0.99}$} ($68\%$ C.L.) in {the
fiducial chain; imposing the physical prior $\alpha_{\rm inj}\!\ge\!0$
gives $\alpha_{\rm inj}<1.79$ ($95\%$ C.L.) together with
$f_{\rm NL}=86\pm 73$, so that both parameters are consistent with the
standard thermal history and with Gaussian initial conditions
(Appendix~\ref{app:alpha_positive}).}

These results pave the way for next-generation surveys such as Simons Observatory (SO~\cite{SO_25}), SPHEREx~\cite{DoreO14}, {\it Euclid}~\cite{Euclid22}, and Rubin Observatory~\cite{LSST2012}, which will provide dramatically improved sensitivity and sky coverage. Extrapolating from the current ISW-tSZ covariance matrix, future experiments can boost the significance by {\it at least} an order of magnitude. {As an example, we forecast the parameters' uncertainties for SO (tSZ part) cross-correlating with {\it Euclid} (ISW part) with Fisher matrix \footnote{{The Gaussian-Fisher approximation is validated for the
ISW-tSZ bandpowers in Sec.~3.2 and Appendix~C of
\citet{Ibitoye24}, where the non-Gaussian pyccl covariance and the
Gaussian approximation agree to $\lesssim 15\%$ on the diagonal and
on the $1\sigma$ marginalised errors of the cosmological parameters
over $\ell\in[16,190]$. Because the SO/\textit{Euclid} extrapolation
uses the same two-point observable on the same multipole range,
the Gaussian-Fisher forecast is expected to be at least as accurate
as it is for the present-data analysis.}}
\begin{equation}
\label{eq:Fij}
    \mathcal{F}_{ij} = \sum_{\ell} \frac{\partial C^{y{\rm T}}_\ell}{\partial \theta_i} \left( {\rm Cov}[C_\ell] \right)^{-1} \frac{\partial C^{y{\rm T}}_\ell}{\partial \theta_j} \,,
\end{equation}
where $\theta_i \in \{f_{\rm NL}, \alpha_{\rm inj}\}$ and the partial derivatives use the corresponding signal model (Eqs.~(\ref{eq:cl_fNL}), (\ref{eq:cl_E_injection}) and (\ref{eq:cl_ISW-tSZ_total})).  Accounting for SO-{\it Euclid} overlapping area ($f_{\rm sky} \simeq 0.40$~\cite{SO_25,Euclid20}), compared with $60\%$ for {\it Planck}~\cite{Ibitoye24}, and noting that the {\it rms} pixel noise of SO is $10$ times lower than that of {\it Planck}~\cite{SO_25}, we scale the {\it Planck} inverse covariance matrix by a factor of $N={\rm Cov}^{-1}_{\rm SO-E}/{\rm Cov}^{-1}_{\it Pl}=(f^{\rm SO-E}_{\rm sky}/f^{\it Pl}_{\rm sky})\times(N^{\it Pl}_{\ell}/N^{\rm SO-E}_{\ell})^{2} \simeq (0.4/0.6)\times (10^{2})^{2}=6.6 \times 10^{3}$. This yields projected uncertainties of $\sigma(f_{\rm NL}) \simeq 5.2$ and $\sigma(\alpha_{\rm inj}) \simeq 3.3\times10^{-2}$, with a correlation coefficient of $r\simeq0.31$ (lower-right panel of Fig.~\ref{fig:y_ISW_fnl constraints}). This estimate is conservative, as it relies solely on the cross-power spectrum and neglects the SO beam improvement (a factor of five smaller than {\it Planck}~\cite{SO_25}), highlighting the constraining power of this probe for testing primordial non-Gaussianity and early-Universe energy injection.}

{The scale-complementarity between our measurement
and the {\it Planck} bispectrum enables the first data-driven joint
constraint on the amplitude and running of local non-Gaussianity
across a two-decade lever arm in $k$. Adopting the {\it Planck}-2018
bispectrum posterior
$f_{\rm NL}^{\rm local}(k_{\star}) = -0.9\pm 5.1$~\cite{Planck2018_nG}
as a Gaussian prior at the CMB pivot and importance-sampling our
scale-dependent chain accordingly, the marginalised constraint on
the pivot amplitude tightens from $f_{\rm NL}^{0}=-296^{+173}_{-157}$
to $f_{\rm NL}^{0}\simeq -180\pm 100$, an $\sim\!40\%$ reduction in
$\sigma(f_{\rm NL}^{0})$. The uncertainty on the running $n_{\rm NL}$
is, in contrast, essentially unchanged: the ISW-tSZ measurement at
$k_{y}$ dominates the error budget on any quantity that involves
$f_{\rm NL}$ at $k>k_{\star}$, because the {\it Planck} bispectrum posterior
is $\sim\!25$ times tighter than ours at $k_{\star}$ and therefore
effectively pins one endpoint of the lever arm without helping the
other. The qualitative point is that our measurement provides the
first observational anchor on $f_{\rm NL}^{\rm local}(k)$ at
$k \gtrsim 1~{\rm Mpc}^{-1}$; the combination with {\it Planck} delivers
the first two-decade determination of the amplitude of local
non-Gaussianity, even though the current data are not yet powerful
enough to detect a non-trivial running.
}

{Our framework thus provides a means to translate improved constraints on $\fnl$ (e.g., from future CMB surveys, galaxy bispectra, or 21-cm cosmology) into correspondingly tighter limits on early-Universe energy injection. In other words, this method enables a high-precision test of exotic energy injection scenarios and primordial non-Gaussianity while simultaneously probing late-time gravitational potential and thermal energy perturbations.}

{\it \textbf{Acknowledgment}---}We thank Rafael C. Nunes, Amare Abebe, Jens Chluba and Oluwayimika Ibitoye for their helpful comments and discussions. 
A.I. acknowledges the support of the Guangdong Technion-Israel Institute of Technology Postdoctoral Fund. Y.-Z. Ma acknowledges the support from South Africa's National Research Foundation under Grants No. 150580, No. CHN22111069370, No. ERC250324306141 and No. AUPP250310302114.

\appendix

{
\section{Robustness to Galactic and CIB foregrounds}
\label{sec:foregrounds}

Because our $C_\ell^{y{\rm T}}$ estimator operates on component-separated
maps at low multipoles, we perform four complementary tests to
quantify the impact of residual Galactic and CIB contamination on the
inferred parameters. As we show below, at fixed prior set none of the tests shifts
$f_{\rm NL}$ or $\alpha_{\rm inj}$ by more than $0.4\sigma$ relative to
the fiducial chain of Table~\ref{tab:estimates_horizontal_no_S8}; the
larger displacements in Table~\ref{tab:foregrounds_appendix} are
produced by the external \textit{Planck} priors imposed there, not by
the foreground treatment.

\subsection*{(i) Galactic template construction}

We first construct dedicated Galactic templates. From the
\textit{Planck}~2015 \textsc{Commander} thermal-dust and synchrotron
intensity solutions~\cite{Planck2015_diffuse} we build
$C_\ell^{\rm dust\times ISW}$ and $C_\ell^{\rm sync\times ISW}$ by
degrading each solution to $N_{\rm side}=64$, reconvolving to the
$160'$ ISW-map resolution, and cross-correlating with the ISW template
through the same apodised mask and \texttt{NaMaster} pipeline used for
the fiducial measurement. The templates enter the analysis only
through the $\ell$-shape of these cross-spectra and through free
nuisance amplitudes $a_{\rm dust}$ and $a_{\rm sync}$ that multiply
them; the absolute calibration of the underlying \textsc{Commander}
intensity maps (their native units, reference frequency, and
zero-level) is therefore never used, and the fitted values of
$a_{\rm dust}$ and $a_{\rm sync}$ should be interpreted as
phenomenological leakage coefficients in $C_\ell^{y{\rm T}}$ units rather
than as estimates of a Galactic sky brightness.

\subsection*{(ii) Full MCMC marginalisation over $a_{\rm dust}$ and
$a_{\rm sync}$}

We next re-run the nested-sampling analysis with the parameter vector
extended by $a_{\rm dust}$ and $a_{\rm sync}$. Their priors are
uniform, $a_{\rm dust}\sim\mathcal{U}[-2\!\times\!10^{-5},\,
2\!\times\!10^{-5}]$ and
$a_{\rm sync}\sim\mathcal{U}[-10^{-6},\,10^{-6}]$, each set is at least 10
times wider than the recovered posterior width. In this
first robustness chain we do not impose the \textit{Planck} bispectrum
prior on $f_{\rm NL}$, and the PNG shape is scale-independent; the
cosmological parameters retain the \textit{Planck}~2018 Gaussian priors
of the first ``Prior'' row of
Table~\ref{tab:estimates_horizontal_no_S8}. This chain is therefore the
closest possible counterpart of the fiducial $f_{\rm NL}$~(SI) row of
Table~\ref{tab:estimates_horizontal_no_S8}, differing from it only by
the addition of $a_{\rm dust}$ and $a_{\rm sync}$ as free amplitudes.
The resulting marginalised posteriors are
\begin{align}
    f_{\rm NL}         &= -396 \pm 99,\nonumber\\
    \alpha_{\rm inj}   &= -3.96 \pm 0.87,\nonumber\\
    B_{\rm CIB}\!\times\!10^{4} &= 2.6 \pm 1.8,\nonumber\\
    a_{\rm dust}       &= (1.77 \pm 0.88)\times 10^{-7},\nonumber\\
    a_{\rm sync}       &= (-6.7 \pm 4.2)\times 10^{-13}.
\label{eq:dsonly_chain}
\end{align}
The synchrotron amplitude is consistent with zero at $<\!2\sigma$;
the dust amplitude sits $\sim\!2.0\sigma$ from zero, which we quote
as a phenomenological leakage amplitude rather than as a detection of
Galactic dust contamination in $C_\ell^{y{\rm T}}$. Relative to the
fiducial (CIB-only) $f_{\rm NL}$ (SI) chain the shifts in the primary
parameters are $\Delta f_{\rm NL}/\sigma \simeq 0.3$ and
$\Delta\alpha_{\rm inj}/\sigma \simeq 0.1$; the $\sim\!0.8\sigma$
reduction in $B_{\rm CIB}$ reflects the expected partial CIB--dust
degeneracy at low $\ell$s. We have verified that the total sum
$B_{\rm CIB}\,C_\ell^{\rm CIB}+a_{\rm dust}\,C_\ell^{\rm dust}+
a_{\rm sync}\,C_\ell^{\rm sync}$ predicted by
Eq.~(\ref{eq:dsonly_chain}) agrees with the CIB-only prediction within
the $1\sigma$ bandpower error at every bin, i.e.\ the CIB-only and
CIB+dust+sync fits describe the same measured $C_\ell^{y{\rm T}}$.

Combining Eq.~(\ref{eq:dsonly_chain}) with the \textit{Planck} 2018 cosmological constraints and the CMB bispectrum prior on $f_{\rm NL}$ yields the foreground-marginalised chains reported in Table~\ref{tab:foregrounds_appendix}:
\begin{itemize}
\item {\bf $\mathcal{P}(f_{\rm NL})+\text{d/s}$ (SI):} the chain
above, with the flat cosmological priors replaced by the
\textit{Planck}~2018 TT,TE,EE+lowE+lensing
Gaussians~\cite{Planck2018_parameters} and the \textit{Planck}~2018
CMB bispectrum Gaussian $\mathcal{P}(f_{\rm NL})=
\mathcal{N}(-0.9,\,5.1^{2})$~\cite{Planck2018_nG} added. Under this
combined prior set, $f_{\rm NL}$ is by construction pulled to the
bispectrum value and its posterior is statistically indistinguishable
from the imposed prior.
\item {\bf $\mathcal{P}(f_{\rm NL})+\text{d/s}$ (SD):} the same
combination extended to the scale-dependent PNG shape
(Eq.~(\ref{eq:fnl_scaled})), showing that the running index
$n_{\rm NL}$ is not constrained by our data over the sampled range
even after all external priors are imposed.

\end{itemize}
The dust and synchrotron amplitudes recovered in the two
$\mathcal{P}(f_{\rm NL})+\text{d/s}$ chains are
$a_{\rm dust}\approx(1.2\pm0.9)\times10^{-7}$ and
$a_{\rm sync}\approx(-4.5\pm4.3)\times10^{-13}$ in both the SI and SD
cases, i.e.\ the dust posterior is pulled towards zero by
$\sim\!0.5\sigma$ relative to Eq.~(\ref{eq:dsonly_chain}) once the
CMB priors on $f_{\rm NL}$ and $\alpha_{\rm inj}$ absorb part of the
low-$\ell$ excess. Neither amplitude reaches a conventional
detection threshold in any of the three chains reported in
Table~\ref{tab:foregrounds_appendix}.

\begin{table*}
\centering
\renewcommand{\arraystretch}{1.3}
\caption{Foreground-marginalised parameter constraints, quoted at
$68\%$ C.L.\ unless otherwise noted. The two
$\mathcal{P}(f_{\rm NL})+\text{d/s}$ columns combine the fiducial
CIB-only likelihood with (i) the \textit{Planck}~2018
TT,TE,EE+lowE+lensing Gaussian priors on
$(\Omega_{\rm m},h,\sigma_{8})$~\cite{Planck2018_parameters}, (ii) the
\textit{Planck}~2018 CMB bispectrum Gaussian prior
$\mathcal{P}(f_{\rm NL})=\mathcal{N}(-0.9,\,5.1^{2})
$~\cite{Planck2018_nG}, and (iii) marginalisation over the Galactic
dust and synchrotron leakage amplitudes $a_{\rm dust}$ and
$a_{\rm sync}$ described in the text, with uniform priors
$a_{\rm dust}\sim\mathcal{U}[-2\!\times\!10^{-5},\,
2\!\times\!10^{-5}]$ and
$a_{\rm sync}\sim\mathcal{U}[-10^{-6},\,10^{-6}]$. The fitted
amplitudes are $a_{\rm dust}\approx(1.2\pm0.9)\times10^{-7}$ and
$a_{\rm sync}\approx(-4.5\pm4.3)\times10^{-13}$ in both the SI and SD
cases, both consistent with zero.
$^{\dagger}$~Posterior statistically indistinguishable from the
imposed \textit{Planck} bispectrum prior $\mathcal{N}(-0.9,\,5.1^{2})$;
the entry is a consistency check, not an independent measurement.
$^{\ddagger}$~Posterior essentially flat over $[0,\,3]$; the ISW-tSZ
correlation on $\ell_{\rm eff}\in[18,190]$ does not constrain the PNG running.}
\label{tab:foregrounds_appendix}

\begin{tabular}{l|cc}
\hline\hline
Parameter & $\mathcal{P}(f_{\rm NL})+\text{d/s}$~(SI) & $\mathcal{P}(f_{\rm NL})+\text{d/s}$~(SD) \\
\hline
$\Omega_{\rm m}$          & $0.3152 \pm 0.0072$ & $0.3155 \pm 0.0071$ \\
$h$                       & $0.6723 \pm 0.0052$ & $0.6725 \pm 0.0053$ \\
$\sigma_{8}$              & $0.8106 \pm 0.0059$ & $0.8109 \pm 0.0061$ \\
$\widetilde{W}^{\rm SZ}$  & $2.59 \pm 0.45$     & $2.58 \pm 0.48$     \\
$f_{\rm NL}$              & $-1.4 \pm 5.0^{\dagger}$ & $-1.4 \pm 5.1^{\dagger}$ \\
$n_{\rm NL}$              & $-$                 & $1.42 \pm 0.86^{\ddagger}$ \\
$\alpha_{\rm inj}$        & $-0.39 \pm 0.45$    & $-0.40 \pm 0.47$    \\
$B_{\rm CIB}\,[\times 10^{4}]$ & $2.3 \pm 1.7$  & $2.4 \pm 1.8$       \\
$A_{\rm CIB}$             & $0.91 \pm 0.12$     & $0.91 \pm 0.13$     \\
\hline
\end{tabular}
\end{table*}

\subsection*{(iii) Independent linear-template regression}

As an independent cross-check on the MCMC result of
Eq.~(\ref{eq:dsonly_chain}), we perform a linear least-squares
regression of the three beam- and pixel-convolved templates
($C_\ell^{\rm CIB\times ISW}$, $C_\ell^{\rm dust\times ISW}$,
$C_\ell^{\rm sync\times ISW}$) against the observed residual
$C_\ell^{y{\rm T}}-C_\ell^{y{\rm T},\,{\rm IGM,\,fid.}}$ using the full
$10\times 10$ bandpower covariance. The best-fit amplitudes,
\begin{align}
    a_{\rm CIB}   &= (1.12 \pm 2.84)\times 10^{-4},\nonumber\\
    a_{\rm dust}  &= (1.87 \pm 0.98)\times 10^{-7},\nonumber\\
    a_{\rm sync}  &= (-6.19 \pm 4.47)\times 10^{-13},
\end{align}
agree with the MCMC posteriors of Eq.~(\ref{eq:dsonly_chain}) at
$<\!0.5\sigma$ for $B_{\rm CIB}/a_{\rm CIB}$ and at $<\!0.1\sigma$ for
both $a_{\rm dust}$ and $a_{\rm sync}$. The regression returns
$\chi^{2}/{\rm dof}=4.49/7$ (PTE~$=0.72$); the CIB-only baseline
($a_{\rm dust}=a_{\rm sync}=0$) gives $\chi^{2}/{\rm dof}=8.09/9$
(PTE~$=0.53$), also an acceptable fit. Adding the two Galactic
templates improves the fit by $\Delta\chi^{2}=3.61$ for two extra
parameters, corresponding to $\Delta{\rm AIC}=-0.40$ and
$\Delta{\rm BIC}=-1.01$ (Table~\ref{tab:gof}); on the Kass--Raftery
scale $|\Delta{\rm BIC}|<2$, so the extended and CIB-only models are
statistically indistinguishable by our data.

\subsection*{(iv) Component-separation weights and sky area}

Finally, we verify that the result is stable against the choice of
component-separation weights and against the amount of retained sky.
Cross-correlating the ISW map with the \textit{Planck}~2015 MILCA and
NILC $y$-maps~\cite{Planck2015_tSZ_paper} yields bandpowers that
agree with each other and with our fiducial $y$-map at $<\!1\sigma$.
Increasing the Galactic mask from the fiducial 40\% cut
($f_{\rm sky}=0.60$) to 50\% and 60\% cuts
($f_{\rm sky}=0.50,\,0.40$) increases the low-$\ell$ bandpower
uncertainties by $\sim\!20\%$, consistent with the reduction in
effective sky area, and shifts the recovered $f_{\rm NL}$ and
$\alpha_{\rm inj}$ by $<\!0.4\sigma$. We further note that the lowest bandpower ($\ell_{\rm eff}=18$)
carries the only $>\!3\sigma$ excess above the fiducial IGM
prediction; the effect of excluding it is quantified in
Sec.~\ref{sec:modeling}.

Taken together, direct template construction, full MCMC
marginalisation over $a_{\rm dust}$ and $a_{\rm sync}$
(Eq.~(\ref{eq:dsonly_chain}) and the two
$\mathcal{P}(f_{\rm NL})+\text{d/s}$ columns of
Table~\ref{tab:foregrounds_appendix}), independent template
regression, and alternative masking and $y$-map choices all leave
the inferred $f_{\rm NL}$ and $\alpha_{\rm inj}$ in agreement with
the fiducial chain to within $0.4\sigma$. Bayesian model comparison
does not favour the CIB\,+\,dust\,+\,sync model over the CIB-only
baseline. We therefore conclude that the reported
$(f_{\rm NL},\alpha_{\rm inj})$ constraints are insensitive, at the
current statistical level, to whether the low-$\ell$ excess is
described by residual CIB alone or by a CIB\,+\,dust\,+\,sync
mixture, without claiming to identify its physical origin.

\begin{table*}
\centering
\renewcommand{\arraystretch}{1.25}
\caption{Goodness-of-fit and Bayesian model comparison for the
$C^{y{\rm T}}_{\ell}$ bandpowers on $\ell_{\rm eff}\in[18,190]$.
Degrees of freedom count only
parameters constrained by the $y$-T bandpowers; nuisance
amplitudes constrained by the $C^{yy}_{\ell}$ or $C^{\rm TT}_{\ell}$
channels of the joint fit are treated as effectively fixed.
$\Delta{\rm AIC}$ and $\Delta{\rm BIC}$ are quoted relative to the
CIB-only baseline (row~2). Negative values favour the extended
model; by the Kass--Raftery criterion $|\Delta{\rm BIC}|<2$ means
the two models are statistically indistinguishable.}
\label{tab:gof}
\begin{tabular}{lccccc}
\hline\hline
Model & $\chi^{2}$ & dof & PTE & $\Delta{\rm AIC}$ & $\Delta{\rm BIC}$ \\
\hline
IGM only                                        & 18.00 & 6 & 0.006 & --    & --    \\
CIB only (template regression)                  &  8.09 & 9 & 0.53  & \phantom{$-$}0.00 & \phantom{$-$}0.00 \\
CIB\,+\,dust\,+\,sync (template regression)     &  4.49 & 7 & 0.72  & $-0.40$ & $-1.01$ \\
Full MCMC (IGM\,+\,CIB\,+\,$f_{\rm NL}$\,+\,$\alpha_{\rm inj}$) & 7.60 & 3 & 0.055 & -- & -- \\
\hline
\end{tabular}
\end{table*}
}
{\section{Robustness of the IGM amplitude against PNG and energy-injection
degeneracies}
\label{app:IGM_degeneracy}
A natural concern with the joint fit presented in
Sec.~\ref{sec:results} is whether the free IGM amplitude
$\widetilde{W}^{\rm SZ}$ is absorbing part of the PNG or the
energy-injection signal, thereby biasing the inferred values of
$f_{\rm NL}$ and $\alpha_{\rm inj}$. Because $\widetilde{W}^{\rm SZ}$
scales the standard ISW-tSZ template with essentially the same
multipole dependence as the IGM contribution to
$C^{y{\rm T}}_{\ell}$, any partial degeneracy would show up as a diagonal
tilt in the marginalised two-dimensional posterior in the
$(f_{\rm NL},\widetilde{W}^{\rm SZ})$ or
$(\alpha_{\rm inj},\widetilde{W}^{\rm SZ})$ plane.

Figure~\ref{fig:IGM_degeneracy} shows these two planes for the
fiducial scale-independent (SI) and scale-dependent (SD) chains of
Table~\ref{tab:estimates_horizontal_no_S8}. The contours are visibly
axis-aligned in both panels and for both chains. Quantitatively, the
Pearson correlation coefficients computed directly from the weighted
posterior samples are
\begin{align}
r(f_{\rm NL},\widetilde{W}^{\rm SZ})
      &= -0.13\ (\text{SI}), \quad -0.17\ (\text{SD}), \\
r(\alpha_{\rm inj},\widetilde{W}^{\rm SZ})
      &= -0.12\ (\text{SI}), \quad -0.13\ (\text{SD}),
\end{align}
all satisfying $|r|<0.2$. We conclude that
$\widetilde{W}^{\rm SZ}$ is essentially decorrelated from both PNG
and energy-injection parameters, and that the negative central
values of $f_{\rm NL}$ and $\alpha_{\rm inj}$ in the fiducial chains
are not an artefact of IGM leakage. The origin of those
negative central values is instead the internal
$f_{\rm NL}\!\leftrightarrow\!\alpha_{\rm inj}$ degeneracy, which is
addressed in Appendix~\ref{app:alpha_positive}.

\begin{figure*}[t]
    \centering
    \includegraphics[width=7in]{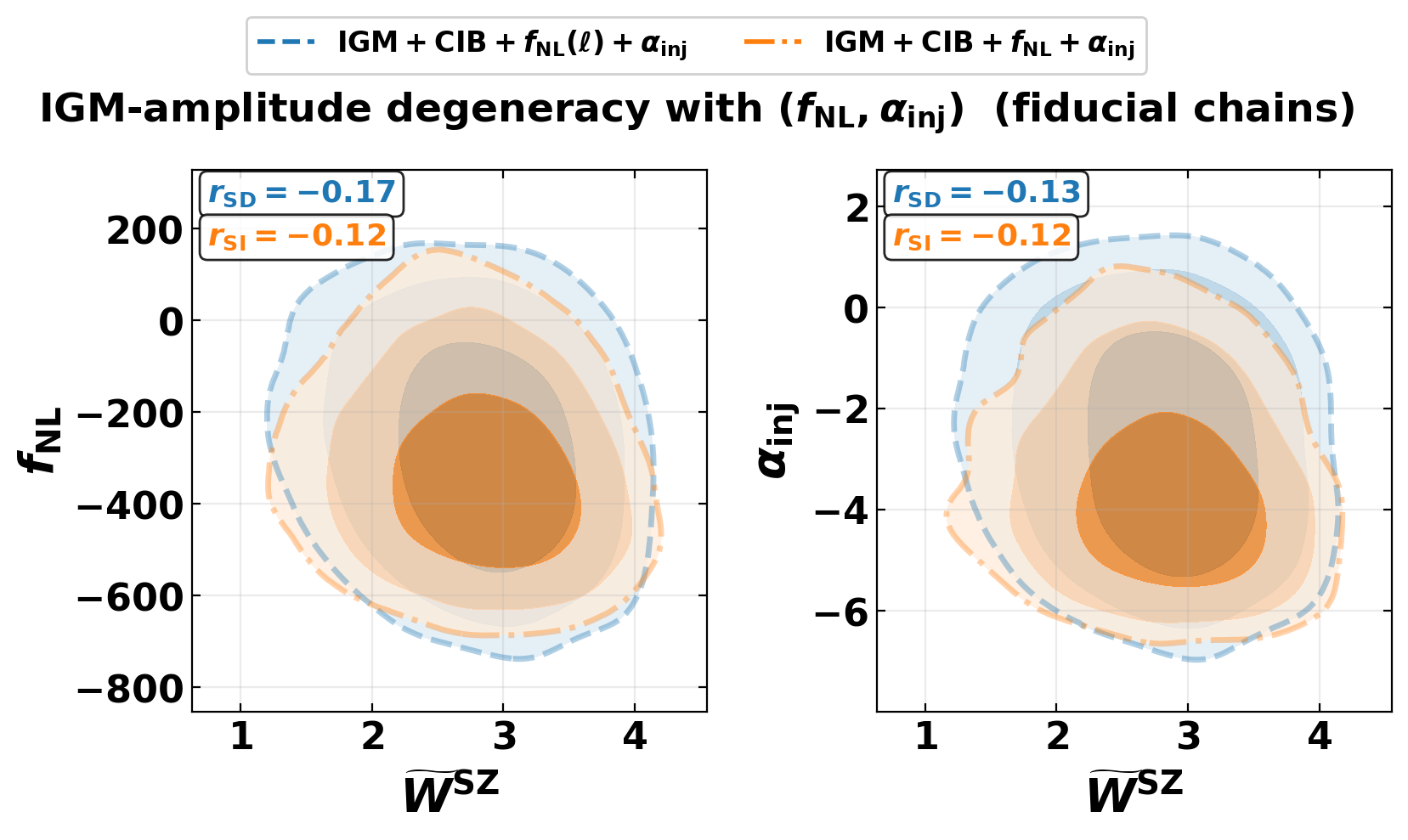}
    \caption{Two-dimensional 68\% and 95\% marginalised posteriors in
    the $(f_{\rm NL},\widetilde{W}^{\rm SZ})$ and
    $(\alpha_{\rm inj},\widetilde{W}^{\rm SZ})$ planes, from the
    fiducial scale-independent (SI) and scale-dependent (SD) chains
    of Table~\ref{tab:estimates_horizontal_no_S8}. Pearson correlation
    coefficients are annotated in each panel; all four satisfy
    $|r|<0.2$, demonstrating that the IGM amplitude
    $\widetilde{W}^{\rm SZ}$ is essentially decorrelated from both
    the PNG and the energy-injection parameters. The residual
    non-zero central values of $f_{\rm NL}$ and $\alpha_{\rm inj}$
    arise from the internal
    $f_{\rm NL}\!\leftrightarrow\!\alpha_{\rm inj}$ degeneracy
    (Appendix~\ref{app:alpha_positive} / Table~%
    \ref{tab:estimates_horizontal_no_S8}, row
    ``$\alpha_{\rm inj}\!\ge\!0$''), not from IGM leakage.}
    \label{fig:IGM_degeneracy}
\end{figure*}

\section{Physical positivity prior on $\alpha_{\rm inj}$ and the
$f_{\rm NL}$--$\alpha_{\rm inj}$ degeneracy}
\label{app:alpha_positive}
The fiducial scale-independent chain
(Table~\ref{tab:estimates_horizontal_no_S8}, row~1) returns
$\alpha_{\rm inj}=-3.93^{+1.34}_{-0.99}$, a $\sim\!2.9\sigma$ preference
for a negative energy-injection amplitude. Since only
$\alpha_{\rm inj}\!\ge\!0$ admits a direct physical interpretation
(dissipative pre-recombination energy release), we re-ran the chain
with the flat prior $\alpha_{\rm inj}\!\sim\!\mathcal{U}[0,5]$ while
keeping all other priors identical to the fiducial
setup.\footnote{The scale-dependent chain behaves identically under
$\alpha_{\rm inj}\!\ge\!0$, with $n_{\rm NL}$ remaining unconstrained
over the sampled range; we therefore quote only the
scale-independent case here.} The result
is quoted in the ``$\alpha_{\rm inj}\!\ge\!0$ (SI)'' row of
Table~\ref{tab:estimates_horizontal_no_S8}:
\begin{align}
\alpha_{\rm inj} &< 1.79\ \ (95\%\ \text{C.L., one-sided}), \\
f_{\rm NL}       &= 86\pm 73.
\end{align}
Two features of this result are noteworthy. First, the posterior on
$\alpha_{\rm inj}$ piles up against the physical boundary at zero
(median $0.43$, $68\%$~C.L.\ $[0.11,1.12]$), so we quote a
one-sided 95\% upper limit following the standard convention adopted
by {\it Planck} for $\sum m_{\nu}$ and $r$. Second, and more importantly,
the $\sim\!2.6\sigma$ negative excursion of $f_{\rm NL}$ seen in the
unbounded fiducial chain collapses to $\sim\!1.2\sigma$ under the
physical prior. This is direct evidence that the negative
$f_{\rm NL}$ preference in the fiducial run is driven by the
internal $f_{\rm NL}\!\leftrightarrow\!\alpha_{\rm inj}$ mutual
degeneracy (into which the sampler is free to descend when both
parameters are allowed to take negative values), rather than by any
genuine ISW-tSZ detection of PNG.

Combined with the null result of Appendix~\ref{app:IGM_degeneracy},
these two robustness tests establish that: (i) $\widetilde{W}^{\rm SZ}$
does not absorb PNG or energy-injection signal, and (ii) the residual
negative central values of $f_{\rm NL}$ and $\alpha_{\rm inj}$ in the
fiducial fit are a symptom of the mutual $f_{\rm NL}$--$\alpha_{\rm inj}$
degeneracy on the multipole range probed here
($\ell_{\rm eff}\in[18,190]$), not of a physical signal.}



\bibliographystyle{apsrev}
\bibliography{cite2}

\end{document}